\documentclass[a4paper]{report}
\usepackage[utf8]{inputenc}
\usepackage[T1]{fontenc}
\usepackage{RJournal}
\usepackage{amsmath,amssymb,array}
\usepackage{booktabs}
\usepackage{hyperref}
\usepackage{verbatim}

\newcommand{\Rn}{{\mathbb{R}}}

\newcommand{\vf}{\boldsymbol{f}}
\newcommand{\vg}{\boldsymbol{g}}

\newcommand{\vr}{\boldsymbol{r}}

\newcommand{\vv}{\boldsymbol{v}}

\newcommand{\vC}{\boldsymbol{C}}

\newcommand{\vI}{\boldsymbol{I}}

\newcommand{\vM}{\boldsymbol{M}}

\newcommand{\vP}{\boldsymbol{P}}

\newcommand{\vR}{\boldsymbol{R}}

\newcommand{\vT}{\boldsymbol{T}}

\newcommand{\vV}{\boldsymbol{V}}

\newcommand{\vX}{\boldsymbol{X}}

\newcommand{\vZ}{\boldsymbol{Z}}

\newcommand{\ind}{1\hspace{-0.7ex}1}

\newcommand{\lan}{ \scriptstyle \mathcal{O}\textstyle}
\newcommand{\N}{{\mathbb{N}}}

\newcommand{\vGamma}{\boldsymbol{\Gamma}}

\newcommand{\vep}{\boldsymbol{\epsilon}}

\newcommand{\vmu}{\boldsymbol{\mu}}

\newcommand{\vSigma}{\boldsymbol{\Sigma}}

\newcommand{\vtheta}{\boldsymbol{\theta}}

\newcommand{\vzeta}{\boldsymbol{\zeta}}

\newcommand{\veins}{{\bf 1}}
\newcommand{\vnull}{{\bf 0}}
\DeclareMathOperator{\Cov}{Cov}
\DeclareMathOperator{\Corr}{Corr}
\DeclareMathOperator{\Var}{Var}
\DeclareMathOperator{\vech}{vech}

\newcommand{\tr}{\operatorname{tr}}

\newcommand{\bqan}{\begin{eqnarray}}
\newcommand{\eqan}{\end{eqnarray}}
\newcommand{\vUpsilon}{\boldsymbol{\Upsilon}}
\newcommand{\vLambda}{\boldsymbol{\Lambda}}
\newcommand{\To}{\longrightarrow}            % Abbildungspfeil
\newcommand{\vepsilon}{\vep}
\newtheorem{Sa}{Theorem}[section]
\newtheorem{theorem}{Theorem}[section]
\newtheorem{re}[Sa]{Remark}

\begin{document}

%% do not edit, for illustration only
\sectionhead{Contributed research article}
\volume{XX}
\volnumber{YY}
\year{20ZZ}
\month{AAAA}

%% replace RJtemplate with your article
\begin{article}
  \title{Testing Hypotheses regarding Covariance and Correlation matrices with the R package CovCorTest}
\author{by Paavo Sattler and Svenja Jedhoff}
\maketitle
\abstract{
In addition to the commonly analysed measures of location, dispersion measurements such as variance and correlation provide many valuable information. Consequently, they play a crucial role in multivariate statistics, which leads to tests regarding covariance and correlation matrices. Furthermore, also the structure of these matrices leads to important hypotheses of interest, since it contains substantial information about the underlying model. In fact, assumptions regarding the structures of covariance and correlation matrices are often fundamental in statistical modelling and testing.
 In this context, semi-parametric settings with minimal distributional assumptions and very general hypotheses are essential for enabling manifold usage. The freely available package \textbf{CovCorTest} provides suitable tests addressing all issues mentioned above, using bootstrap and similar techniques to achieve good performance, particularly in small samples. Additionally, the package offers flexible specification options for the hypotheses under investigation in two central tests, accommodating users with varying levels of expertise, which results in high flexibility and user-friendliness at the same time. This paper also presents the application of \textbf{CovCorTest} for various issues, illustrated by multiple examples, where the tests are applied to a real-world dataset.
}
 \section{Introduction}
 
 Tests for measures of location are widely used in the analysis of multivariate data, such as those arising from repeated measures designs or multiple measurements on the same subject. While expectation vectors are a classical choice for such analyses (\cite{konietschke2020}), alternative parameters such as quantiles (\cite{baumeister2024}), relative effects (\cite{rubarth2022} or \cite{dobler2020a}) and others have been proposed. 
 In any case, measures of dispersion contain substantial information and are, therefore, of central importance in multivariate settings. Especially, covariance and correlation matrices are excellent parameters for analyses, which is, among other reasons, because:
 \begin{itemize}
 \item They are comparably easy to define and to understand.
 \item They qualify the relationships between the components, which most other parameters do not take into account.
 \item In statistical modelling, covariance and correlation matrices and their structures (e.g. autoregressive or compound symmetry) are essential for understanding and describing the underlying processes.
 \item Conditions on the covariance or correlation matrix, such as homogeneity or a special structure/pattern, are among the most frequent assumptions of statistical methods.
 \item Finite second moments and therefore, the existence of variances is usually presumed and sufficient to ensure consistency of estimators for the unknown covariance or correlation matrices.
 \end{itemize}
 Due to their importance, a variety of statistical procedures have been developed to test hypotheses concerning covariance matrices. Unfortunately, most existing methods rely on restrictive distributional assumptions, such as normality in \cite{box1953} and \cite{zhong2017}, or specific conditions on the characteristic function in \cite{gupta2006}, which limit their applicability. Furthermore, several approaches are limited to specific settings, for example, two group settings as in \cite{sakaori2002} or \cite{omelka2012}. Moreover, they typically allow the investigation of only one or two concrete hypotheses, such as equality of covariance matrices (e.g., \cite{yang2012} or \cite{zhu2002}) or equality of two correlation matrices (e.g., \cite{jennrich1970}). This also holds for the software packages implementing these procedures, such as the R-packages \textbf{psych}(\cite{Revelle2019}) or \textbf{CovTestR}(\cite{Barnard2018}).
 
In contrast, the framework introduced in \cite{sattler2022} allows a broad class of hypotheses, building on methodologies familiar from the analysis of location parameters.
At the same time, the less restrictive semi-parametric setting used therein allows broad applicability.
Thereto, test statistics are constructed as quadratic forms and Monte-Carlo techniques and resampling approaches are used to obtain the required quantiles and achieve good small sample performance. This approach was further expanded in \cite{sattler2023} to quite general hypotheses regarding correlation matrices, where the focus was on the Anova-type statistic, and additionally, a Taylor-based Monte-Carlo approach was incorporated. The corresponding R package \textbf{CovCorTest} (\cite{covcortest}) includes the adequate tests from both papers to provide consistent procedures for investigating covariance and correlation matrices, and moreover also combined tests for equality of covariance and correlation matrices at the same time.
Additionally, specific structures of covariance and correlation matrices, such as the Toeplitz form, are of great interest, as they provide meaningful insights into the underlying model. Besides, specific structures are usual model assumptions, but tests regarding these are often insufficient.
 Most existing approaches are limited to specific structures such as compound symmetry (\cite{winer1991} or \cite{votaw1948}) or first-order autoregressive (\cite{mckeown1996}) or focus exclusively on either covariance or correlation matrices (\cite{steiger1980}).
Conversely, the recent work by \cite{sattler2024a} introduces a unified framework for testing a large class of structural hypotheses, known as the linear covariance model, respectively the linear correlation model, and even further structures. These tests, which are related to the approaches presented in \cite{sattler2022} and \cite{sattler2023}, have also been integrated into the package to enable comprehensive and coherent testing of various aspects of covariance and correlation matrices in one place.\\
To facilitate access, the package \textbf{CovCorTest} is available for free from the  Comprehensive R Archive Network (CRAN) at
\url{https://cran.r-project.org/web/packages=CovCorTest}.\\
It contains five main functions (\verb|test_covariance|, \verb|test_correlation|,
\verb|test_covariance_structure|,\\
\verb|test_correlation_structure| and \verb|test_combined|), which are as analogous as possible to make it more user-friendly. This uniform design allows researchers to apply the functions in analogous situations, targeting different aspects of the data. Moreover, this approach allows for the investigation of all related hypotheses within a single package, while the coherent construction and methodology enhance the comparability of the results.
Furthermore, to support users with varying levels of experience, the covariance and correlation tests allow not only the selection of predefined hypotheses, but also the specification of custom hypotheses by providing a hypothesis matrix and the corresponding vector.
Since for the remaining tests, either only one hypothesis is possible (\verb|test_combined|) or only predefined structures can be tested, no such differentiation is necessary there.

Because all underlying approaches and tests are based on empirical covariance and correlation matrices, our functions should only be used for metric data, although further assumptions, such as a repeated-measures design, are not required.

In the subsequent section, we will introduce the statistical multivariate model, the investigated null hypotheses, and the underlying assumptions. Afterwards, we present the implemented inference procedures, followed by the applications from the R package \textbf{CovCorTest} illustrated by some examples using a real-world dataset. After all, a conclusion and outlook complete the paper.

 \section{Statistical Model and Hypotheses}
 We consider a general multigroup setting  with $a\in \N$ groups,
 where each group $i=1,...,a$ contains $n_i\in \N$ observations.
 The total number of observations is denoted by $N = \sum_{i=1}^a n_i$, allowing for potentially unbalanced designs in which group sizes may differ.
 
Then, a general semi-parametric model is used, defined by independent $d$-dimensional random vectors 
\[\vX_{ik}= \vmu_i + \vep_{ik},\] 
where $\vmu_i\in \R^d$  represents the group-specific mean vector, and $\vep_{ik}\in \R^d $  denotes the random deviation for subject $k=1,...,n_i$ from the $i$-th group.
 Here, splitting the indices allows considering factorial designs and longitudinal data, as demonstrated, e.g., in \cite{pauly2015}.
We assume the residuals $\vep_{i1},\dots,\vep_{in_i}$ to be centred $\E(\vep_{i1}) = \vnull_d$ and i.i.d. within each group, while across different groups only independence is needed, but differing distributions are possible.
Further conditions for $i=1,...,a$ are 
\begin{itemize}
\vspace*{0.2cm}

  \item[(A1)] $\frac{n_i}{N}\to \kappa_i\in (0,1]$, for $\min(n_1,\dots,n_a)\to\infty$ 
  \item[(A2)]$\E(||\vep_{i1}||^4) < \infty$
  \item[(A3)]$\Cov(\vep_{i1})=\vV_i=(V_{ijk})_{j,k\leq d}\geq 0$ %\textcolor{red}{überlegen ob notwendig}
  \item[(A4)]$\Var(\vep_{i1\ell})=V_{i\ell\ell}>0$, ~$\ell= 1,...,d$
\end{itemize}
where  (A4) is only required for procedures involving correlation matrices since it guarantees the existence of $\vR_i=\Corr(\vep_{i1})$.
Additionally, since correlations are only meaningful for at least two components, we assume $d\geq 2$.

Since both covariance and correlation matrices are symmetric, it is natural to consider an appropriate vectorisation to transform these matrices into vectors containing the same information as the original matrix. For the covariance matrices, the so-called half-vectorisation operation $\vech$ is used, which extracts the upper-triangular (including diagonal) elements of a $d\times d$ matrix, yielding a $p=d(d+1)/2$ dimensional vector. The upper-half-vectorisation $\vech^-$, on the contrary, only takes the values above the diagonal into a $p_{-}=d(d-1)/2$ dimensional vector, which is useful for correlation matrices whose diagonal entries are always one and hence contain no information. Now, based on hypothesis matrices $\vC_{\vv}\in \Rn^{m_1 \times p}$ respectively
$\vC_{\vr}\in \Rn^{m_2 \times p_{-}}$ and vectors $\vzeta_{\vv}\in \Rn^{m_1}$, $ \vzeta_{\vr}\in \Rn^{m_2}$ general linear hypotheses can be formulated. These hypotheses are defined based on the pooled covariance vector $\vv= (\vech(\vV_1)^\top,\dots,\vech(\vV_a)^\top)^\top$  respectively the pooled correlation vector $\vr= (\vech^{-}(\vR_1)^\top,\dots,\vech^{-}(\vR_a)^\top)^\top$ as 
\begin{equation}\label{Hypothesen}
     \mathcal{H}_0^{\vv}:\vC_{\vv}\vv=\vzeta_{\vv}\quad\text{ and}\quad \mathcal{H}_0^{\vr}:\vC_{\vr}\vr=\vzeta_{\vr}.\end{equation}
% For the corresponding structures, the model is a bit more general, with
 %\[\mathcal{H}_0^{\vv,s}:\widetilde\vC_{\vv}\vf(\vv)=\widetilde\vzeta_{\vv}\quad\text{ and}\quad \mathcal{H}_0^{\vr,s}:\widetilde\vC_{\vr}\vg(\vr)=\widetilde\vzeta_{\vr},\]
%with adequate functions $\vf:\Rn^{p}\mapsto \Rn^{\ell_1}, \vg:\Rn^{p_{i}}\mapsto \Rn^{\ell_2}$ matrices $\widetilde\vC_{\vv}\in \Rn^{m_1 \times \ell_1},\widetilde\vC_{\vr}\in \Rn^{m_2 \times \ell_2}$ and vectors $\widetilde\vzeta_{\vv}\in \Rn^{m_1},\widetilde\vzeta_{\vr}\in \Rn^{m_2}$.
Popular hypotheses included in this model, among many others, are
\begin{itemize}
    \item Equal covariance matrices: $\vV_1=...=\vV_a$
    \item Equal traces of covariance matrices: $\tr(\vV_1)=...=\tr(\vV_a)$
    \item Equal equal variances $(\vV_1)_{11}=...=(\vV_1)_{dd}$
    
    \item Equal correlation matrices: $\vR_1=...=\vR_a$
    \item Equal correlated components: $\vech^-(\vR_1)_{1}=...=\vech^-(\vR_1)_{p_{-}}$
    \item Uncorrelated components: $\vech^-(\vR_1)=\vnull_{p_{-}}$
    \item Trace of the covariance matrix has a special value $\gamma\in \Rn_+$: $\tr(\vV_1)=\gamma$
    \item Specific covariance matrix $\vV\in\Rn^{d\times d}$: $\vV_1=\vV$
\end{itemize}
Note that the same hypothesis can often be expressed in multiple ways through different choices of $\vC$ and $\vzeta$; see \cite{sattler2023a, sattler2025} for a detailed discussion. In the case of $\vzeta=\vnull$, unique projection matrices exist to check the underlying hypotheses, while in general, the considered hypothesis matrices do not have to be square, idempotent or symmetrical. Based for example on the centering matrix $\vP_d=\vI_d-\veins_{d\times d}/d$ and the Kronecker product, this leads to
 \[(\vP_a \otimes \vI_p)\vv=\vnull_{ap} \quad \Leftrightarrow \quad \vV_1=...=\vV_a\quad \text{resp.} \quad (\vP_a \otimes \vI_{p_{-}})\vr=\vnull_{ap_{-}} \quad \Leftrightarrow \quad \vR_1=...=\vR_a\]
 or  
 \[\vP_{p_{-}}\vr=\vnull_{p_{-}} \quad \Leftrightarrow \quad \vech^-(\vR_1)_{1}=...=\vech^-(\vR_1)_{p_{-}}\]
and similar for the other hypotheses.\\

In addition to these general hypotheses, structural hypotheses on the form of the covariance or correlation matrices are of substantial interest. For covariance matrices, common structures include
\begin{itemize}
    \item  Compound symmetry

 \item Diagonality

 \item First-order autoregressive (AR(1))

 \item Sphericity

 \item Toeplitz structure,
\end{itemize}
see \cite{kincaid2005} for more details regarding these structures.

For correlation matrices, analogous—but typically fewer—structures exist,  which often arise as heterogeneous versions of other structures such as heterogeneous autoregressive, heterogeneous compound symmetry or heterogeneous Toeplitz.\\
Except for the autoregressive structure and all its versions, all of these covariance and correlation structures can be expressed within the general hypothesis framework using appropriate choices of $\vC$ and $\vzeta$, as shown in \cite{sattler2024a}. 
{Therein they introduced the linear covariance structure model, comparable to  \cite{anderson1973} or \cite{szatrowski1980}, given for $q<p$  through}
\begin{align}
 \label{eq:lcsmodel}
  \mathcal{V}=\{\vV(\vtheta)\in \R^{d\times d}:\vV(\vtheta)=\vV_0+\theta_1 \vV_1+...+\theta_q \vV_q, \ \vtheta = ( \theta_1, \dots, \theta_q)^\top\in \R^q\} \cap \mathcal{COV}_{d \times d}
\end{align}
{where $\vV_0,...,\vV_q$ are known linearly independent symmetric matrices with $\vV_1,...,\vV_q\neq\vnull_{d\times d}$ and $\mathcal{COV}_{d \times d} \subset \R^{d \times d}$ is the cone of covariance matrices, i.e., symmetric positive semi-definite matrices. In an analogous way, also the linear correlation structure model was defined. A central result thereof is that each covariance matrix structure from one of these models can be expressed through $\mathcal{H}_0^{\vv}:\vC_{\vv}\vv=\vzeta_{\vv}$ respectively $\mathcal{H}_0^{\vr}:\vC_{\vr}\vr=\vzeta_{\vr}$. Although these expressions are not unique, the package includes a function \verb|get_hypothesis|  which for given $\vech(\vV_0),(\vech(\vV_1),...,\vech(\vV_q))$ resp. $\vech^-(\vR_0),(\vech^-(\vR_1),...,\vech^-(\vR_q))$ (for the linear correlation structure model) returns a possible hypothesis matrix $\vC$ and corresponding vector $\vzeta$.}

This highlights the versatility and generality of the presented model in capturing also a wide range of relevant covariance and correlation structures.

 \section{Inference}
To evaluate the hypotheses formulated in the previous section, we develop inference procedures based on the asymptotic distribution of vectorised covariance and correlation estimates.

For our hypotheses regarding covariance matrices, we consider the vectorised empirical group covariances $\widehat \vv_i:=\vech\left((n_i-1)^{-1}\sum_{i=1}^{n_i}(\vX_{ik}-\overline \vX_{i})(\vX_{ik}-\overline \vX_{i})^\top)\right)$ where $\overline \vX_{i}$ is the sample mean of group $i$.
Stacking these vectors across all $a$ groups yields the pooled vector $\widehat\vv=(\widehat\vv_1^\top,...,\widehat\vv_a^\top)^\top$.
 Then, for developing a statistical test, we use that under assumption (A2) and (A3) for $i=1,\dots, a$, the group-wise covariance vectors satisfy \[\sqrt{N} (\widehat \vv_i -\vv_i)\stackrel{\mathcal {D}}{\longrightarrow} {\mathcal{N}_{p}\left(\vnull_{p},\kappa_i^{-1}\cdot \vSigma_i\right)}\] with $\vSigma_i=\Cov(\vech(\vepsilon_{i1}\vepsilon_{i1}^\top))$.  Consequently, under the null hypothesis $\mathcal{H}_0^{\vv}$ and (A1) it holds
\[\sqrt{N} (\vC_{\vv}\widehat \vv -\vzeta_{\vv})\stackrel{\mathcal {D}}{\longrightarrow} {\mathcal{N}_{m_1}\left(\vnull_{m_1},\vC_{\vv}\vSigma \vC_{\vv}^\top\right)},\]
with the block-diagonal covariance matrix $\vSigma=\bigoplus_{i=1}^a {\kappa_i}^{-1}\cdot  \vSigma_i$.\\

In the same way, $\widehat\vr_i$ denotes the empirical counterpart of the upper-half-vectorised correlation matrix of group $i$ and $\widehat\vr$ is the corresponding pooled vector.
 Following the results in \cite{sattler2023} (where also the exact form can be found), there exists a continuous matrix-valued function $\vM\in \Rn^{p_-\times p}$  such that
 \begin{equation}\label{relation} \sqrt{n_i}(\widehat \vr_i-\vr_i)=\vM(\vv_i,\vr_i)\cdot \sqrt{n_i} (\widehat \vv_i-\vv_i)+\lan_P(1). \end{equation}
Using this result and the previous asymptotic expansion, we obtain the joint asymptotic distribution under $\mathcal{H}_0^{\vr}$ as
\[\sqrt{N}(\vC_{\vr}\widehat \vr-\vzeta_{\vr})\stackrel{\mathcal D}{\To}\mathcal{N}_{m_2}\left(\vnull_{m_2},\vC_{\vr}\bigoplus\nolimits_{i=1}^a {\kappa_i}^{-1}\vUpsilon_i\vC_{\vr}^\top\right)=\mathcal{N}_{m_2}(\vnull_{m_2},\vC_{\vr}\vUpsilon\vC_{\vr}^\top), \]
where we use $\vUpsilon_i:=\vM(\vv_i,\vr_i)\vSigma_i\vM(\vv_i,\vr_i)^\top$.

Based on these limit distributions, to test our multivariate hypotheses, we use the ATS (Anova-type-statistic),  initially developed in \cite{brunner1997}, which is given in this case through:

\begin{alignat}{2}
ATS_{\vv}&={N}\left[\vC_{\vv}(\widehat\vv-\vzeta_{\vv})\right]^\top \left[\vC_{\vv}(\widehat\vv-\vzeta_{\vv})\right]/\tr(\vC_{\vv} \widehat{\vSigma}\vC_{\vv}^\top),\label{ATS1}\\
ATS_{\vr}&={N}\left[\vC_{\vr}(\widehat\vr-\vzeta_{\vr})\right]^\top \left[\vC_{\vr}(\widehat\vr-\vzeta_{\vr})\right]/\tr(\vC_{\vr} \widehat{\vUpsilon}\vC_{\vr}^\top)\label{ATS2}.
\end{alignat}
Here for the unknown covariance, we use the estimator $\widehat \vSigma: =\bigoplus_{i=1}^a {N}/{n_i}  \cdot \widehat \vSigma_i$ , where
$$\widehat \vSigma_i=\frac{1}{n_i-1} \sum\limits_{k=1}^{n_i}\left[\vech\left(\widetilde \vX_{ik}\widetilde \vX_{ik}^\top- \sum\limits_{\ell=1}^{n_i}\frac{\widetilde \vX_{i\ell}\widetilde \vX_{i\ell}^\top}{n_i}\right)\right]\left[\vech\left(\widetilde \vX_{ik}\widetilde \vX_{ik}^\top- \sum\limits_{\ell=1}^{n_i}\frac{\widetilde \vX_{i\ell}\widetilde \vX_{i\ell}^\top}{n_i}\right)\right]^\top.
$$ is based on $\widetilde \vX_{ik}:=\vX_{ik}-\overline \vX_{i\cdot}$, the centered version of observation $k$ in group $i$. Based on this, for the correlation we get a plug-in estimator through \[\widehat \vUpsilon:=\bigoplus_{i=1}^a \frac{n_i}{N}\widehat \vUpsilon_i,\quad\widehat \vUpsilon_i=\vM(\widehat\vv_i,\widehat\vr_i)\widehat \vSigma_i\vM(\widehat\vv_i,\widehat\vr_i)^\top.\]
There also exist alternative versions of the ATS, which, for example, allow the approximation through an F-distribution (see e.g. \cite{brunner2019} or \cite{thiel2024}). However, for the statistics defined in   \eqref{ATS1} and \eqref{ATS2}, the asymptotic null distribution is known, enabling valid inference without stringent distributional assumptions.

\begin{theorem}\label{Verteilung}  
\phantom{1}\begin{itemize}
\item[a)]
Under Assumption (A1)-(A3) and the null hypothesis $\mathcal{H}_0^v:\vC_{\vv}\vv=\vzeta_{\vv}$, the $ATS_{\vv}$ defined by \eqref{ATS1}
 has, asymptotically, a ``weighted $\chi^2$-distribution''. That is, % we have
$$ATS_{\vv}\stackrel{\mathcal {D}}{\To} \sum_{\ell=1}^{m_1} \lambda_\ell B_\ell,
$$
where $B_\ell \stackrel{i.i.d.} {\sim} \chi_1^2$ and $\lambda_\ell, \ell =1,\dots, m_1,$ are the eigenvalues of $(\vC_{\vv} \vSigma\vC_{\vv}^\top)/\tr(\vC_{\vv} {\vSigma}\vC_{\vv}^\top)$.

\item[b)]
Under Assumption (A1)-(A4) and the null hypothesis $\mathcal{H}_0^r:\vC_{\vr}\vr=\vzeta_{\vr}$, the $ATS_{\vr}$ defined by \eqref{ATS2}
 has, asymptotically, a ``weighted $\chi^2$-distribution''. That is, % we have
$$ATS_{\vr}\stackrel{\mathcal {D}}{\To} \sum_{\ell=1}^{m_2} \lambda_\ell B_\ell,
$$
where $B_\ell \stackrel{i.i.d.} {\sim} \chi_1^2$ and $\lambda_\ell, \ell =1,\dots, m_2,$ are the eigenvalues of $(\vC_{\vr} \vUpsilon\vC_{\vr}^\top)/\tr(\vC_{\vr} {\vUpsilon}\vC_{\vr}^\top)$.
\end{itemize}
\end{theorem}

A Monte-Carlo technique is one useful approach to obtain the required quantiles of these weighted sums for this statistic. For the covariance matrix, with our consistent covariance estimator $\widehat \vSigma$, we get estimators $\hat \lambda_1,...,\hat \lambda_{m_1}$ for the unknown eigenvalues. With these weights we  generate appropriate realisations  $B_1,...,B_{m_2} \stackrel{i.i.d.} {\sim} \chi_1^2$. These realisations and our estimators allow calculating a weighted sum $\sum\limits_{\ell=1}^{m_2} \hat \lambda_\ell B_\ell$, which converges in distribution against the above limit distribution. Repeating this procedure many times allows to calculate the required Monte-Carlo quantiles, and with an analogous procedure for the correlation for $\alpha \in (0,1)$, we get asymptotic level $\alpha$ tests
through
 
 \[\begin{array}{lrl}
&\varphi_{\vv}^{MC}&=\ind\left(ATS_{\vv}>q_{\vv,1-\alpha}^{MC}\right) ,\\
&\varphi_{\vr}^{MC}&=\ind\left(ATS_{\vr}> q_{\vr,1-\alpha}^{MC}\right).
\end{array}\]\\

Another technique for obtaining quantiles that requires more computational effort but usually yields better small-sample performance is a bootstrap approach. For hypotheses regarding covariance matrices, the parametric bootstrap mimics the limit distribution and afterwards the structure of the test statistic by generating bootstrap vectors $\vZ_{i 1}^\star,...,\vZ_{i n_i}^\star\stackrel{i.i.d.}{\sim} \mathcal N_{p}\left(\vnull_p,\widehat{\vSigma}_{i}\right),$ based on given realisations $\vX_{i 1},..., \vX_{i n_i}$ with estimators $\widehat \vSigma_{i}$. Then 
$\widehat \vSigma_i^\star$, the empirical covariance matrix of the bootstrap sample $\vZ_{i1}^\star,...,\vZ_{in_i}^\star$   and $\widehat \vSigma^\star:=\bigoplus_{i=1}^a{N}/{n_i}\cdot  \widehat \vSigma_i^\star$ are calculated. Following the bootstrap realization $ATS_{\vv}^\star=N[ \vC\overline \vZ^\star  ]^\top[ \vC\ \overline \vZ^\star  ]\big/\tr(\vC \widehat \vSigma^\star \vC^\top)$ is calculated,
using the bootstrap means $\overline \vZ_i^\star=n_i^{-1}\sum_{k=1}^{n_i}\vZ_ik^\star$ and $\overline \vZ^\star=({\overline \vZ_1^\star}^\top,...,{\overline \vZ_a^\star}^\top)^\top $.
Then, based on a large number of repetitions, the bootstrap quantile $q_{\vv,1-\alpha}^{\star}$ is determined.

In a similar way, by replacing $\widehat \vSigma$ and $\widehat \vSigma_i^\star$ through $\widehat \vUpsilon$ and $\widehat \vUpsilon_i^\star$ the bootstrap quantile for the correlation  $q_{\vr,1-\alpha}^{\star}$ can be determined. This allows formulating asymptotic bootstrap tests for the respective hypothesis and $\alpha \in (0,1)$ by

 \[\begin{array}{lrl}
&\varphi_{\vv}^{\star}&=\ind\left(ATS_{\vv}>q_{\vv,1-\alpha}^{\star}\right) ,\\
&\varphi_{\vr}^{\star}&=\ind\left(ATS_{\vr}> q_{\vr,1-\alpha}^{\star}\right).
\end{array}\]

For the investigation of hypotheses regarding correlations, an additional approach exists to obtain the required quantiles. It is called a Taylor-based Monte-Carlo technique and can be seen as a mixture between Monte-Carlo and bootstrap, using a refined version of the expansion in equation \eqref{relation} to simulate the asymptotic distribution directly. Further details are provided in \cite{sattler2023}.\\
\begin{re}
Despite its high flexibility, the linear hypotheses introduced in \eqref{Hypothesen} did not allow the formulation of all structures for covariance and correlation matrices. In particular, the autoregressive structure and all of its versions include some proportionality between the matrix components that cannot be captured through linear constraints on the vectorised covariance or correlation matrices.
To accommodate such nonlinear structures, we consider transformations of the parameter vectors.
Therefore, a kind of differentiable transformation $\vf$ of the vectorised covariance resp. $\vg$ of the correlation is required. Therewith, the hypothesis can be formulated through $\mathcal{H}_0^{\vv,s}:\widetilde\vC_{\vv}\vf(\vv)=\widetilde\vzeta_{\vv}$ or $\mathcal{H}_0^{\vr,s}:\widetilde\vC_{\vr}\vg(\vr)=\widetilde\vzeta_{\vr}$. By replacing the vectorised covariance matrix respectively correlation matrix by their transformed pendant, and adding the corresponding Jacobian matrix, the above-mentioned tests can be adapted. A detailed discussion, along with a useful transformation, referred to as the subdiagonal-mean-ratio, is provided in \cite{sattler2024a}.
\end{re}

Although these tests are very versatile through the less restrictive assumptions and the general class of hypotheses, they do not by themselves yield insight into the specific reason a hypothesis is rejected — e.g., whether the differences arise due to variance differences, dependence structure, or both.
%But, when comparing multiple groups, we are often interested in whether they differ regarding their variances or their dependency structures.
Since correlation matrices reflect the dependency structure but contain no information about the components' variances, and covariance matrices contain no standardised dependency measure, neither is the best solution on its own. To address this issue, \cite{sattler2023} proposes a combined test for simultaneously assessing group differences in both covariance and correlation. This approach is based on the concept of Multiple Contrast Testing (see, e.g., \cite{bretz2001}) and is applicable when comparing two groups ($a=2$).
In case of rejection of equal covariance matrices, it also provides information on whether the dependency, the variance, or both are responsible for it.
This test is based on a vector-value test statistic

\[\vT=\begin{pmatrix}
T_1\\
\vdots\\
T_p
\end{pmatrix}= \sqrt{N}\left(\begin{pmatrix}
\widehat V_{1 11}\\
\widehat V_{1 22}\\
\vdots\\
\widehat V_{1 dd}\\
\widehat \vr_1
\end{pmatrix}- \begin{pmatrix}
\widehat V_{2 11}\\
\widehat V_{2 22}\\
\vdots\\
\widehat V_{2 dd}\\
\widehat \vr_2
\end{pmatrix}\right)\]
for which it holds under the $\mathcal{H}_0^{\vv}:\vV_1=\vV_2$ the convergence $\vT\stackrel{\mathcal{D}}{\to}\mathcal{N}_p(\vnull_p,\vGamma)$, while the explicit form of the asymptotic covariance matrix  $\vGamma$ can be found in \cite{sattler2023}.

To calculate the required quantiles to use this test statistic, a simulation-based method similar to the Taylor-based Monte Carlo approach is employed to mimic the limit distribution. To this aim for both groups $i=1,2$ independent realisations $\vLambda_i\sim\mathcal{N}_{p}(\vnull_{p},\widehat \vSigma_i)$ are generated, and subsequently transformed to $\vLambda_i^{Tay}$. The extensive formula for the used transformation and its derivation is also part of the online supplement in \cite{sattler2023}. From this, we compute $\vT^{1,Tay}=\sqrt{N}(n_1^{-1/2}\vLambda_1^{Tay}-n_{2}^{-1/2}\vLambda_2^{Tay})$ and repeat this $B\in \N$ times, allowing to derive empirical $\beta$ quantiles for each component, denoted as $q_{\ell,\beta}$. Subsequently, the local $\beta$ has to be determined in a way that it controls the family-wise type-I-error rate by $\alpha$, which is assured through the approach of \cite{munko2024} by using

\[{\widetilde \beta= \max\left(\beta \in \mathcal{B}\Big\lvert \sum\limits_{b=1}^B \max\limits_{\ell=1,...,p}\frac{\left(1-\ind\left(q_{\ell,\frac{\beta}{2}}\leq T_\ell^{b,Tay}\leq q_{\ell,1-\frac{\beta}{2}}\right)\right)}{B}\leq \alpha\right)}\] 
with $\mathcal{B}=\left\lbrace 0,\frac{1}{B},..., \frac{B-1}{B}\right\rbrace$.

An asymptotic level-$\alpha$ test of the null hypothesis of equal covariance matrices is then defined through
\[\max\left[\max\limits_{\ell=1,...,d}\left(1-\ind\left(q_{\ell, \frac{\widetilde\beta}{2}}\leq T_\ell\leq q_{\ell,1-\frac{\widetilde\beta}{2}}\right)\right),\max\limits_{\ell=d+1,...,p}\left(1-\ind\left(q_{\ell, \frac{\widetilde\beta}{2}}\leq T_\ell\leq q_{\ell,1-\frac{\widetilde\beta}{2}}\right)\right)\right],\]
while the arguments resulting in rejection contain further information. If the first argument is equal to one, this means that the variances differ, while a second argument of one implies that the correlation differs. This also leads to three p-values: one for the variances, one for the correlation, and one global p-value, which is the maximum of the other two.
In contrast to all other considered tests, this combined test of variance and correlation is only defined for $a=2$ groups.

\section{Software and examples}
In this section, we will analyse datasets regarding different hypotheses to illustrate the usage of the testing procedures implemented in the \textbf{CovCorTest} package. Thereto, first, the syntax of our functions is explained, which enables their application to a real-world dataset. Since most of the implemented tests follow a common construction, their application is consistent and intuitive. Therefore, we begin with a general overview of the syntax before highlighting specific differences. Hereby, especially \verb|test_combined| differs through its more specific hypothesis and setting, resulting in fewer options.
\subsection{Syntax}
\begin{itemize}
    \item \textit{data}: The dataset for which the test should be conducted, while two different input formats are possible. One possibility is a list, where each list element is a matrix containing the values from one group as columns. Since each column is an observation, all matrices must have the same number of rows, while the group sample size and, therefore, the number of columns can differ.
    The second format are all groups together in one large matrix, where the observations are ordered by the group. In this case, an accompanying vector \textit{nv} has to be provided, containing the group sample size, while for single-group tests, this value is not necessary.
    Since the structure tests of covariance or correlation matrices are only applicable for one group, the second format is mandatory in this case.
    \item \textit{hypothesis:}
    \begin{itemize}
           \item For increased user-friendliness, the functions \verb|test_covariance|, \verb|test_correlation|,\\
\verb|test_covariance_structure| and \verb|test_correlation_structure| allow to choose from a list of predefined hypotheses,
     which depends on the considered test, respectively function:\vspace{0.05cm}\\
     \underline{Covariance matrix:}
        Here, in the case of one group, the options are "equal" (denoting equal covariances between the components), "given-trace", "given-matrix" and "uncorrelated". In the case of multiple groups, the predefined hypotheses are "equal" (now denoting equal covariance matrices in the different groups),  "equal-trace" and "equal-diagonals".\\
        \underline{Correlation matrix:} For one group, the hypotheses are "equal-correlated", denoting equal correlation between the components and "uncorrelated", which means that all components have a correlation of zero. In the case of multiple groups, the only hypothesis is "equal-correlated", which here denotes the equality of correlation matrices.\\
        \underline{Covariance Structure:} For covariance structures, the selectable hypotheses are autoregressive ("autoregressive" or "ar"), first-order autoregressive  ("fo-autoregressive" or "fo-ar"), diagonal ("diagonal" or "diag"), sphericity ("sphericity" or "spher"), compound symmetry ("compoundsymmetry" or "cs") and toeplitz ("toeplitz" or "toep").\\
        \underline{Correlation Structure:} The predefined structures of the correlation matrices are heterogeneous autoregressive ("hautoregressive" or "har"), heterogeneous toeplitz ("htoeplitz" or "htoep"), heterogeneous compound symmetry ("hcompoundsymmetry", "hcs") and diagonal ("diagonal" or "diag").
        \item For \verb|test_covariance| and \verb|test_correlation|, more experienced users could specify custom hypotheses through a hypothesis matrix together with a corresponding vector, as introduced in \cite{sattler2022} or \cite{sattler2023}, especially for less standard hypotheses, which increase flexibility. Both refer to the corresponding kind of upper triangular vectorisation, and unlike in other procedures, the hypothesis matrix here has neither to be a projection matrix nor square. { This especially allows structures from the linear covariance structure model and the linear correlation structure model by using the function} \verb|get_hypothesis|.
     \item The function \verb|test_combined|, which jointly assesses the equality of covariance and correlation matrix, is restricted to one specific hypothesis. Consequently, this argument is not available in that context.
    \end{itemize}

    \item \textit{method:}
    Specifies the method by which the quantile for the test is calculated. With \textit{BT}, the quantiles are determined through a parametric bootstrap and with \textit{MC}  through a Monte-Carlo technique, both based on \cite{sattler2022} and  \cite{sattler2023}. In the latter, there is even another method, a Taylor-based Monte-Carlo approach, which leads to the option \textit{Tay} for  \verb|test_correlation|. Again, the combined test works only for a Taylor-based Monte-Carlo approach, which makes any options here unnecessary.
    \item \textit{repetitions}: The number of Monte Carlo or bootstrap repetitions used in the approximation of the quantiles. This number substantially influences both the test's performance as well as its computation time. Since a small number of repetitions increases the influence of the randomness on the results, a predefined value of \textit{1000} is a good trade-off and substantially smaller numbers are not recommended.
   % \item \textit{seed}: Since all procedures contain an incidental portion, a seed can be used to maintain repeatability, where the default setting is no seed. This seed is only used for this function, and afterwards, it is reset.
\end{itemize}

\subsection{Output}
Here, we must differentiate between the combined test and all other implemented test functions.

For the latter ones, the output consists of different parts, which are mostly general and contains more descriptive information about the conducted test, like the investigated null hypothesis (through the label from the list or the hypothesis matrix $\vC$ and the corresponding vector $\vzeta$), the used method for calculating the critical values, the used number of repetitions and the vector containing the group sample sizes. Moreover, the estimated covariance matrix for the test statistic, as well as the value of the test statistic and the p-value, are returned. The print function only gives a short and compact summary of the conducted test, containing the number of groups, the tested hypothesis, the resampling method together with the used number of repetitions, and especially the value of the test statistic and p-value.

For the combined test, the hypothesis, the number of groups, and the resampling method are fixed, since only one hypothesis can be tested for two groups. Moreover, based on the used Taylor-based Monte-Carlo approach, no covariance estimator of the test statistic is used and, therefore, also not returned. Moreover, the test statistic here is vector-valued and non-standardised, making the components hardly comparable. Since conclusions based on the value of the test statistic are usually not possible here, these values are not printed to avoid misinterpretations. 
However, this test results in not only one p-value but also in three of them. A \verb|$pvalue_Variances| for the local hypothesis of equal variances, a \verb|$pvalue_Correlations| for the local hypothesis of equal correlations, and \verb|$pvalue_Total| for the global hypothesis of equal covariance matrices (and therefore both, equal correlations and equal variances). All of them are returned and printed through the function.

\subsection{Examples}
To demonstrate the usage of the main functions and to illustrate typical outputs, we present several example analyses based on real-world data.
For this purpose, we use the dataset EEG, which concludes multivariate measurements from a neurological study conducted at the University Clinic of Salzburg, Department of Neurology. This dataset is part of the \textsc{R}-package \textit{manova.rm} (\cite{Friedrich2019}), and contains $N=160$ patients from both genders, with three different diagnoses of impairments (subjective cognitive complaints (SCC), mild cognitive impairment (MCI), and Alzheimer's disease (AD)). Thereby, two neurological parameters (z-score of the brain rate and Hjorth complexity) were measured at three different locations of the head: temporal, frontal, and central. This design results in a multivariate factorial structure with multiple response variables and multiple independent groups and was further described in \cite{bathke2018}. 
 In the following sections, we apply the core testing procedures from \textbf{CovCorTest} to this dataset, showcasing the practical implementation and interpretation of results.

\subsubsection{Covariance matrix test}
Due to the sample sizes, we will check whether there is homogeneity of covariance matrices for women with these diagnoses, leading to $a=3$ unbalanced groups. The test is performed twice, using the Monte-Carlo approach and the parametric bootstrap as a comparison. For the latter, both kinds of specifying the hypothesis of interest are used. 
\begin{verbatim}
> data("EEGwide", package = "MANOVA.RM")
> vars <- colnames(EEGwide)[1:6]
> SCCFemale <- t(EEGwide[EEGwide$sex == "F" & EEGwide$diagnosis == "SCC",vars])
> MCIFemale <- t(EEGwide[EEGwide$sex == "F" & EEGwide$diagnosis == "MCI",vars])
> ADFemale <- t(EEGwide[EEGwide$sex == "F" & EEGwide$diagnosis == "AD",vars])

> X <- list(SCCFemale, MCIFemale, ADFemale)
> nv <- unlist(lapply(X, ncol))

> set.seed(123)                
> test_covariance(X = X, nv = nv, hypothesis = "equal", method = "MC", repetitions = 1000)

     	 Covariance Test 
 	    3  groups

 Hypothesis: 		equal
 Teststatistic value: 	2.6042
 p-value: 	 	p = 0.024
 
 p-value computed using Monte-Carlo-technique with B=1000 repetitions
 

> set.seed(123)                
> test_covariance(X = X, nv = nv, hypothesis = "equal", method = "BT", repetitions = 1000)

       	 Covariance Test 
 	        3  groups

 Hypothesis: 		equal
 Teststatistic value: 	2.6042
 p-value: 	 	p = 0.023
 
 p-value computed using Bootstrap with B=1000 repetitions
 

> P3 <- diag(1,3,3) - 1\3
> Hypothesismatrix <- P3%x%diag(1, 6, 6)
> Hypothesisvector <- matrix(0, 18, 1)
> set.seed(123)
> test_covariance(X = X, nv = nv, C = Hypothesismatrix, Xi = Hypothesisvector,
                  method = "BT", repetitions = 1000)

       	 Covariance Test 
 	        3  groups

 Hypothesis: 		equal
 Teststatistic value: 	2.6042
 p-value: 	 	p = 0.023
 
 p-value computed using Bootstrap with B=1000 repetitions

\end{verbatim}

As expected, since the value of the test statistic is not affected by the resampling method, both values coincide, while we get slightly different p-values. With p-values of  0.024 resp 0.023, the null hypothesis of equal covariance matrices in all groups would be rejected at the usual level of $5\%$.

%\textcolor{red}{Beispiel-Code für Cov-Gleichheit für 2 Gruppen. Am besten mit allen Methoden, ein 1 Gruppen verfahren ist ja bei der Struktur dran}
\subsubsection{Correlation matrix test}
After comparing the covariance matrices for women with these diagnoses, the correlation matrices are also investigated for equality. In addition to the parametric bootstrap, the test is also conducted for the Taylor-based Monte-Carlo approach, specifically developed for testing vectorised correlation matrices.
\begin{verbatim}

> set.seed(123)                
> test_correlation(X = X, nv = nv, hypothesis = "equal-correlated", method = "BT",
                   repetitions = 1000)

       	 Correlation Test 
 	    3  groups

 Hypothesis: 		equal-correlated
 Teststatistic value: 	0.7432
 p-value: 	 	p = 0.602
 
 p-value computed using Bootstrap with B=1000 repetitions

> set.seed(123)
> test_correlation(X = X, nv = nv, hypothesis = "equal-correlated", method = "Tay",
                   repetitions = 1000)

       	 Correlation Test 
 	    3  groups

 Hypothesis: 		equal-correlated
 Teststatistic value: 	0.7432
 p-value: 	 	p = 0.596
 
 p-value computed using Taylor-based Monte-Carlo-approach with B=1000 repetitions
\end{verbatim}

Here, the p-values are way higher than for comparison of covariance matrices, such that this hypothesis is not rejected for usual $\alpha$ levels. This is not surprising, since the hypothesis of equal correlation matrices is a subset of the hypothesis of equal covariance matrices.

\subsubsection{Combined test}
Instead of testing the equality of covariance matrices and the equality of correlation matrices for two groups separately, the combined test from \cite{sattler2023} allows us to do both at the same time. This allows for the null hypothesis of equal covariance matrices to determine whether the difference lies in the correlations and, therefore, the dependency, the variances, or even both.

\begin{verbatim}
> data("EEGwide", package = "MANOVA.RM")
> vars <- colnames(EEGwide)[1:6]
> X <- list(t(EEGwide[EEGwide$sex == "M" & EEGwide$diagnosis == "AD",vars]),
                t(EEGwide[EEGwide$sex == "M" & EEGwide$diagnosis == "MCI",vars]))
> nv <- unlist(lapply(X, ncol))

> set.seed(123)
> test_combined(X, nv)

       	 Combined Test 

 p-value-Variances: 	 	     p = 0.418
 p-value-Correlations: 	 	  p = 0.016
 p-value-Total: 	 	         p = 0.016
 
 p-values computed using Taylor-based Monte-Carlo-approach with B=1000 repetitions 
\end{verbatim}

From the resulting three p-values, we see that both covariance matrices differ, while only a difference in the correlation can be determined, and it was not possible to prove a difference in the variances.

\subsubsection{Structure tests}
In this analysis, we want to investigate whether the position of the measuring points influences the measured values. If we reject the hypothesis of the covariance matrix being a compound symmetry matrix, it is verified that there is an effect of the position. Moreover, to check whether the correlations between the measure points are also equal, we additionally check for a heterogeneous compound symmetry structure by using the correlation matrix.
\begin{verbatim}
> X <- as.matrix(EEGwide[EEGwide$sex == "W" & EEGwide$diagnosis == "AD",
     c("brainrate_temporal", "brainrate_frontal","brainrate_central",
     "complexity_temporal","complexity_frontal", "complexity_central")])

> set.seed(123)   
> test_covariance_structure(X = X, structure = "compoundsymmetry", method = "MC")

       	 Covariance Test 
 	    one group

 Hypothesis: 		compoundsymmetry
 Teststatistic value: 	3.055
 p-value: 	 	p = 0.026
 
 p-value computed using Monte-Carlo-technique with B=1000 repetitions

> set.seed(123)
> test_correlation_structure(X = X, structure = "hcompoundsymmetry", method = "MC")

 	 Correlation Test 
 	    one group

 Hypothesis: 		hcompoundsymmetry
 Teststatistic value: 	5.5229
 p-value: 	 	p < 1e-3
 
 p-value computed using Monte-Carlo-technique with B=1000 repetitions
\end{verbatim}
Since both structures are rejected at a usual significance level of $5\%$, not only was the influence of the measure points' location verified, but also that the correlation between them differs.\\\\
We can also test the compound symmetry structure of the covariance matrix, using a hypothesis formulation based on \verb!get_hypothesis! and \verb!TestCovariance!.\\

\begin{verbatim}
> X <- t(as.matrix(EEGwide[EEGwide$sex == "W" & EEGwide$diagnosis == "AD",
         c("brainrate_temporal", "brainrate_frontal","brainrate_central",
         "complexity_temporal","complexity_frontal", "complexity_central")]))
 
> v0=rep(0,21)
> vauxilariy=c(1,rep(0,5),1,rep(0,4),1,rep(0,3),1,rep(0,2),1,0,1)
> V=cbind(vauxilariy,1-vauxilariy)
> Hypothesis=get_hypothesis(v0,V)
> set.seed(123)   
> test_covariance(X = X,C = Hypothesis$hypothesis_matrix, Xi = Hypothesis$hypothesis_vector, 
                  method = "MC", repetitions = 1000)

         Covariance Test 
            one group

 Hypothesis:            C * v = Xi
 Teststatistic value:   4.0926
 p-value:               p = 0.006
 
 p-value computed using Monte-Carlo-technique with B=1000 repetitions 
\end{verbatim}

Although this is the same test procedure, data set, and seed, the p-value differs. This indicates that for the ATS, the test decision typically depends on the representation of the null hypothesis, as also described in \cite{sattler2023a}.

 \section{Conclusion and Outlook}
In the presented paper, we illustrate the functionality and scope of the R package \textbf{CovCorTest}, which provides a comprehensive framework for hypothesis testing concerning covariance and correlation matrices in multivariate data analysis.
It contains tests for both parameters and different hypotheses in a uniform way, which excels through its applicability to a large class of possible hypotheses and fewer distribution assumptions. Apart from multi-group settings, single groups are also allowed, as well as factorial designs.
In the main, it contains functions for five different tests: regarding covariance matrices, correlation matrices, a combined test for equality of covariance and correlation matrices, 
and tests regarding their corresponding structures.
To this end, quantiles, based on various techniques including resampling and Monte Carlo approaches, are used. Although the Monte-Carlo approach leads to the least time-consuming procedure, a parametric bootstrap for covariances and a Taylor-based Monte-Carlo approach for correlations are preferable for smaller sample sizes. 
 
To include actual developments and procedures in this area, we aim to update this package on a regular basis. Planned future developments include further hypotheses (especially additional structures and patterns) and new test statistics, such as multiple contrast tests for hypotheses regarding covariance and correlation matrices. Also, the combined correlation and covariance test could be expanded to accommodate more than two groups, and tests for multiple structures at once, as mentioned in \cite{sattler2024a}, could be incorporated. Further, the output of a confidence ellipsoid is planned as a further extension, together with the corresponding plot. Also, a graphical user interface (GUI) is possible to simplify the usage, especially for more applied users. Finally, the results from \cite{sattler2025} could be realised in the future, to reduce calculation effort.
 This would allow a higher number of bootstrap and Monte-Carlo repetitions and thereby increase the performance for large datasets.

 \section{Acknowledgement}
Svenja Jedhoff gratefully acknowledges the support from TRR 391 Spatio-temporal Statistics for the Transition of Energy and Transport (520388526), funded by the Deutsche Forschungsgemeinschaft (DFG, German Research Foundation).

\bibliographystyle{apalike}
\bibliographystyle{unsrtnat}

%\phantomsection
\addcontentsline{toc}{chapter}{Bibliography}
\bibliography{Literaturnew}

\begin{thebibliography}{35}
\providecommand{\natexlab}[1]{#1}
\providecommand{\url}[1]{\texttt{#1}}
\expandafter\ifx\csname urlstyle\endcsname\relax
  \providecommand{\doi}[1]{doi: #1}\else
  \providecommand{\doi}{doi: \begingroup \urlstyle{rm}\Url}\fi

\bibitem[Anderson(1973)]{anderson1973}
T.~W. Anderson.
\newblock {Asymptotically Efficient Estimation of Covariance Matrices with
  Linear Structure}.
\newblock \emph{The Annals of Statistics}, 1\penalty0 (1):\penalty0 135 -- 141,
  1973.
\newblock \doi{10.1214/aos/1193342389}.
\newblock URL \url{https://doi.org/10.1214/aos/1193342389}.

\bibitem[Barnard and Young(2018)]{Barnard2018}
B.~Barnard and D.~Young.
\newblock \emph{Testing functions for Covariance Matrices}, 2018.
\newblock URL \url{https://covtestr.bearstatistics.com/}.

\bibitem[Bathke et~al.(2018)Bathke, Friedrich, Pauly, Konietschke, Staffen,
  Strobl, and Höller]{bathke2018}
A.~C. Bathke, S.~Friedrich, M.~Pauly, F.~Konietschke, W.~Staffen, N.~Strobl,
  and Y.~Höller.
\newblock Testing mean differences among groups: Multivariate and repeated
  measures analysis with minimal assumptions.
\newblock \emph{Multivariate Behavioral Research}, 53:\penalty0 348--359, 03
  2018.
\newblock \doi{10.1080/00273171.2018.1446320}.

\bibitem[Baumeister et~al.(2024)Baumeister, Ditzhaus, and
  Pauly]{baumeister2024}
M.~Baumeister, M.~Ditzhaus, and M.~Pauly.
\newblock Quantile-based manova: A new tool for inferring multivariate data in
  factorial designs.
\newblock \emph{Journal of Multivariate Analysis}, 199:\penalty0 105246, 2024.
\newblock ISSN 0047-259X.
\newblock \doi{10.1016/j.jmva.2023.105246}.
\newblock URL
  \url{https://www.sciencedirect.com/science/article/pii/S0047259X23000921}.

\bibitem[Box(1953)]{box1953}
G.~E.~P. Box.
\newblock Non-normality and tests on variances.
\newblock \emph{Biometrika}, 40\penalty0 (3-4):\penalty0 318--335, 1953.
\newblock \doi{10.1093/biomet/40.3-4.318}.

\bibitem[Bretz et~al.(2001)Bretz, Genz, and Hothorn]{bretz2001}
F.~Bretz, A.~Genz, and L.~A. Hothorn.
\newblock On the numerical availability of multiple comparison procedures.
\newblock \emph{Biometrical Journal}, 43\penalty0 (5):\penalty0 645--656, 2001.
\newblock
  \doi{https://doi.org/10.1002/1521-4036(200109)43:5<645::AID-BIMJ645>3.0.CO;2-F}.

\bibitem[Brunner et~al.(1997)Brunner, Dette, and Munk]{brunner1997}
E.~Brunner, H.~Dette, and A.~Munk.
\newblock Box-type approximations in nonparametric factorial designs.
\newblock \emph{Journal of the American Statistical Association}, 92\penalty0
  (440):\penalty0 1494--1502, 1997.
\newblock \doi{10.2307/2965420}.

\bibitem[Brunner et~al.(2019)Brunner, Bathke, and Konietschke]{brunner2019}
E.~Brunner, A.~C. Bathke, and F.~Konietschke.
\newblock \emph{Rank and Pseudo-Rank Procedures for Independent Observations in
  Factorial Designs}.
\newblock Springer, 2019.

\bibitem[Dobler et~al.(2020)Dobler, Friedrich, and Pauly]{dobler2020a}
D.~Dobler, S.~Friedrich, and M.~Pauly.
\newblock Nonparametric manova in meaningful effects.
\newblock \emph{Annals of the Institute of Statistical Mathematics},
  72\penalty0 (5):\penalty0 997--1022, 2020.
\newblock \doi{10.1007/s10463-019-00717-3}.
\newblock URL \url{https://doi.org/10.1007/s10463-019-00717-3}.

\bibitem[Friedrich et~al.(2019)Friedrich, Konietschke, and
  Pauly]{Friedrich2019}
S.~Friedrich, F.~Konietschke, and M.~Pauly.
\newblock \emph{MANOVA.RM: Analysis of Multivariate Data and Repeated Measures
  Designs}, 2019.
\newblock URL \url{https://cran.r-project.org/package=MANOVA.RM}.
\newblock R package version 0.3.2.

\bibitem[Gupta and Xu(2006)]{gupta2006}
A.~K. Gupta and J.~Xu.
\newblock On some tests of the covariance matrix under general conditions.
\newblock \emph{Annals of the Institute of Statistical Mathematics},
  58\penalty0 (1):\penalty0 101--114, Mar 2006.
\newblock ISSN 1572-9052.
\newblock \doi{10.1007/s10463-005-0010-z}.

\bibitem[Jennrich(1970)]{jennrich1970}
R.~I. Jennrich.
\newblock An asymptotic chi square test for the equality of two correlation
  matrices.
\newblock \emph{Journal of the American Statistical Association}, 65\penalty0
  (330):\penalty0 904--912, 1970.
\newblock \doi{10.1080/01621459.1970.10481133}.

\bibitem[Kincaid(2005)]{kincaid2005}
C.~Kincaid.
\newblock Guidelines for selecting the covariance structure in mixed model
  analysis.
\newblock \emph{Statistics and Data Analysis}, 30, 01 2005.

\bibitem[Konietschke et~al.(2020)Konietschke, Schwab, and
  Pauly]{konietschke2020}
F.~Konietschke, K.~Schwab, and M.~Pauly.
\newblock Small sample sizes: A big data problem in high-dimensional data
  analysis.
\newblock \emph{Statistical Methods in Medical Research}, 30\penalty0
  (3):\penalty0 687--701, 2020.
\newblock \doi{10.1177/0962280220970228}.
\newblock URL \url{https://doi.org/10.1177/0962280220970228}.
\newblock Veröffentlicht am 24. November 2020.

\bibitem[McKeown and Johnson(1996)]{mckeown1996}
S.~P. McKeown and W.~D. Johnson.
\newblock Testing for autocorrelation and equality of covariance matrices.
\newblock \emph{Biometrics}, 52\penalty0 (3):\penalty0 1087--1095, 1996.
\newblock ISSN 0006341X, 15410420.
\newblock \doi{10.2307/2533070}.
\newblock URL \url{http://www.jstor.org/stable/2533070}.

\bibitem[Munko et~al.(2024)Munko, Ditzhaus, Dobler, and Genuneit]{munko2024}
M.~Munko, M.~Ditzhaus, D.~Dobler, and J.~Genuneit.
\newblock Rmst-based multiple contrast tests in general factorial designs.
\newblock \emph{Statistics in Medicine}, 43\penalty0 (10):\penalty0 1849--1866,
  2024.
\newblock \doi{10.1002/sim.10017}.

\bibitem[Omelka and Pauly(2012)]{omelka2012}
M.~Omelka and M.~Pauly.
\newblock Testing equality of correlation coefficients in two populations via
  permutation methods.
\newblock \emph{Journal of Statistical Planning and Inference}, 142\penalty0
  (6):\penalty0 1396--1406, 2012.
\newblock \doi{10.1016/j.jspi.2011.12.018}.

\bibitem[Pauly et~al.(2015)Pauly, Ellenberger, and Brunner]{pauly2015}
M.~Pauly, D.~Ellenberger, and E.~Brunner.
\newblock Analysis of high-dimensional one group repeated measures designs.
\newblock \emph{Statistics}, 49\penalty0 (6):\penalty0 1243–1261, July 2015.
\newblock ISSN 1029-4910.
\newblock \doi{10.1080/02331888.2015.1050022}.

\bibitem[Revelle(2019)]{Revelle2019}
W.~Revelle.
\newblock \emph{psych: Procedures for Psychological, Psychometric, and
  Personality Research}.
\newblock Northwestern University, Evanston, Illinois, 2019.
\newblock URL \url{https://CRAN.R-project.org/package=psych}.
\newblock R package version 1.9.12.

\bibitem[Rubarth et~al.(2022)Rubarth, Sattler, Zimmermann, and
  Konietschke]{rubarth2022}
K.~Rubarth, P.~Sattler, H.~G. Zimmermann, and F.~Konietschke.
\newblock Estimation and testing of wilcoxon-mann-whitney effects in factorial
  clustered data designs.
\newblock \emph{Symmetry}, 14\penalty0 (2), 2022.
\newblock ISSN 2073-8994.
\newblock \doi{10.3390/sym14020244}.
\newblock URL \url{https://www.mdpi.com/2073-8994/14/2/244}.

\bibitem[Sakaori(2002)]{sakaori2002}
F.~Sakaori.
\newblock Permutation test for equality of correlation coefficients in two
  populations.
\newblock \emph{Communications in Statistics-simulation and Computation},
  31:\penalty0 641--651, 01 2002.
\newblock \doi{10.1081/SAC-120004317}.

\bibitem[Sattler and Dobler(2024)]{sattler2024a}
P.~Sattler and D.~Dobler.
\newblock Testing for patterns and structures in covariance and correlation
  matrices.
\newblock 2024.
\newblock \doi{10.48550/arxiv.2310.11799}.

\bibitem[Sattler and Pauly(2023)]{sattler2023}
P.~Sattler and M.~Pauly.
\newblock Testing hypotheses about correlation matrices in general manova
  designs.
\newblock \emph{TEST}, 33:\penalty0 496--516, Dec. 2023.
\newblock \doi{10.1007/s11749-023-00906-6}.

\bibitem[Sattler and Rosenbaum(2025)]{sattler2025}
P.~Sattler and M.~Rosenbaum.
\newblock Choice of the hypothesis matrix for using the anova-type-statistic.
\newblock \emph{Statistics \& Probability Letters}, page 110356, 2025.
\newblock ISSN 0167-7152.
\newblock \doi{10.1016/j.spl.2025.110356}.
\newblock URL
  \url{https://www.sciencedirect.com/science/article/pii/S0167715225000021}.

\bibitem[Sattler and Zimmermann(2023)]{sattler2023a}
P.~Sattler and G.~Zimmermann.
\newblock Choice of the hypothesis matrix for using the wald-type-statistic.
\newblock \emph{Statistics \& Probability Letters}, page 110038, 2023.
\newblock ISSN 0167-7152.
\newblock \doi{10.1016/j.spl.2024.110038}.
\newblock URL
  \url{https://www.sciencedirect.com/science/article/pii/S0167715224000075}.

\bibitem[Sattler et~al.(2022)Sattler, Bathke, and Pauly]{sattler2022}
P.~Sattler, A.~C. Bathke, and M.~Pauly.
\newblock Testing hypotheses about covariance matrices in general manova
  designs.
\newblock \emph{Journal of Statistical Planning and Inference}, 219:\penalty0
  134--146, 2022.
\newblock ISSN 0378-3758.
\newblock \doi{10.1016/j.jspi.2021.12.001}.
\newblock URL
  \url{https://www.sciencedirect.com/science/article/pii/S0378375821001269}.

\bibitem[Sattler et~al.(2025)Sattler, Jedhoff, Pauly, Bathke, and
  Dobler]{covcortest}
P.~Sattler, S.~Jedhoff, M.~Pauly, A.~C. Bathke, and D.~Dobler.
\newblock \emph{CovCorTest: Statistical Tests for Covariance and Correlation
  Matrices and their Structures}, 2025.
\newblock URL \url{https://CRAN.R-project.org/package=CovCorTest}.
\newblock R package version 1.1.0.

\bibitem[Steiger(1980)]{steiger1980}
J.~H. Steiger.
\newblock Testing pattern hypotheses on correlation matrices: Alternative
  statistics and some empirical results.
\newblock \emph{Multivariate Behavioral Research}, 15\penalty0 (3):\penalty0
  335--352, 1980.
\newblock \doi{10.1207/s15327906mbr1503_7}.
\newblock PMID: 26794186.

\bibitem[Szatrowski(1980)]{szatrowski1980}
T.~H. Szatrowski.
\newblock Necessary and sufficient conditions for explicit solutions in the
  multivariate normal estimation problem for patterned means and covariances.
\newblock \emph{The Annals of Statistics}, 8\penalty0 (4):\penalty0 802--810,
  1980.
\newblock ISSN 00905364, 21688966.
\newblock URL \url{http://www.jstor.org/stable/2240767}.

\bibitem[Thiel et~al.(2024)Thiel, Sattler, Bathke, and Zimmermann]{thiel2024}
K.~E. Thiel, P.~Sattler, A.~C. Bathke, and G.~Zimmermann.
\newblock Resampling nancova: Nonparametric analysis of covariance in small
  samples.
\newblock 2024.

\bibitem[Votaw(1948)]{votaw1948}
D.~F. Votaw.
\newblock Testing compound symmetry in a normal multivariate distribution.
\newblock \emph{The Annals of Mathematical Statistics}, 19\penalty0
  (4):\penalty0 447--473, 1948.
\newblock ISSN 0003-4851.
\newblock \doi{10.1214/aoms/1177730145}.
\newblock URL \url{http://www.jstor.org/stable/2236016}.

\bibitem[Winer et~al.(1991)Winer, Brown, and Michels]{winer1991}
B.~Winer, D.~R. Brown, and K.~M. Michels.
\newblock \emph{Statistical principles in experimental design}.
\newblock McGraw-Hill, New York, 1991.
\newblock \doi{10.2307/3172747}.

\bibitem[Yang and DeGruttola(2012)]{yang2012}
Y.~Yang and V.~DeGruttola.
\newblock Resampling-based methods in single and multiple testing for equality
  of covariance correlation matrices.
\newblock \emph{The international journal of biostatistics}, 8:\penalty0
  Article 13, 01 2012.
\newblock \doi{10.1515/1557-4679.1388}.

\bibitem[Zhong et~al.(2017)Zhong, Lan, Song, and Tsai]{zhong2017}
P.-S. Zhong, W.~Lan, P.~X.~K. Song, and C.-L. Tsai.
\newblock Tests for covariance structures with high-dimensional repeated
  measurements.
\newblock \emph{Ann. Statist.}, 45\penalty0 (3):\penalty0 1185--1213, 2017.
\newblock \doi{10.1214/16-AOS1481}.

\bibitem[Zhu et~al.(2002)Zhu, Ng, and Jing]{zhu2002}
L.-X. Zhu, K.~Ng, and P.~Jing.
\newblock Resampling methods for homogeneity tests of covariance matrices.
\newblock \emph{Statistica Sinica}, 12:\penalty0 769--783, 07 2002.

\end{thebibliography}
\address{Paavo Sattler\\
  TU Dortmund University, Department of Statistics\\
  44221 Dortmund\\
 Germany\\}
\email{paavo.sattler@tu-dortmund.de}

\address{Svenja Jedhoff\\
  TU Dortmund University, Department of Statistics\\
  44221 Dortmund
\\
 Germany\\}
\email{svenja.jedhoff@tu-dortmund.de}
\end{article}

\end{document}